# Dynamic Model for Formation of Twinned Martensite Crystals

M.P. Kashchenko and V.G. Chashchina

Physics Chair, Ural State Forest Engineering University, Sybirskiy trakt, 37,
620100, Ekaterinburg, Russia mpk46@mail.ru

Abstract

Based on original investigations into the stages of nucleation and growth of martensite, a consideration has been given to dynamic models of formation of martensite plates with a fine structure of transformation twins, which are compatible with the supersonic growth rate of martensite crystals. Along with relatively long quasi-longitudinal waves, which determine the orientation of the habit plane, the control wave process includes relatively short longitudinal waves, which act in synchronism and control the growth of the main component of a regular twin structure. Preference is given to a model containing, at the initial moment of time, the only active dynamic cell capable of periodic reproduction in the interphase region at the stage of the martensite crystal growth. Conditions providing the ratio between the components of the twin structure, which would be optimal for minimization of the transformation time, have been discussed. A transition to final deformations, which are two to three orders of magnitude larger than the threshold deformations, have been made. The calculated macroscopic morphological attributes, which were deduced from dynamic considerations, have been compared with the experimental data and crystallogeometrical calculation.

## 1 Introduction

Reconstructive martensitic transformations are commonly characterized by the formation of fine structure of transformation twins. In addition to discussing a range of general problems associated with the behavior of martensitic transformations, here we address a still other important application of the dynamic theory of crystal growth developed in [1, 2], i.e., the process of twinning under γ - α martensitic transformations in iron-based alloys in crystals with the habitus planes of the $\{3\ 9\ 11\}_\gamma \div \{3\ 10\ 15\}_\gamma$ type. In this case one observes twin structure formation (see e.g., [3, 4]), which involves alternate layers of the master and twinning components possessing (in the parent phase) a $<100>_\gamma$ or $<010>_\gamma$ orientations of the main compressive axes of Bain's deformation Both components of the twin structure have a common tensile axis with the direction close to $<001>_\gamma$. The ratio of the volume fractions β of twinning components lies approximately within the interval 1<β<2, their distribution is regular, and their combination along the $\{110\}_\gamma$ planes transiting into the $\{112\}_\alpha$ planes is taken to be nearly coherent.

The problem of a twinning mechanism is critical for the development of the dynamic theory of martensitic transformations. As far back as 1982 in [5], the author put forward a hypothesis according to which a control over twin structure formation should be interpreted as a consequence of consistent propagation of comparatively short- and long-wave atomic displacements. Pairs of displacement waves (for convenience termed here as long-wave or $l$ –waves) responsible for formation of a habitus plane possess wavelengths on the order of the martensitic plane thickness, $\lambda_\ell \sim (0.1 \div 1) \mu m$. Pairs of longitudinal waves (for convenience termed here short-wave or s–waves) travelling along the orthogonal directions $<100>_\gamma$ and $<010>_\gamma$, prescribe the boundaries $\{110\}_\gamma$ (corresponding to the twin boundaries $\{112\}_\alpha$) and are characterized by the wavelengths $\lambda_s \sim 10^{-2} \mu m$ (about the thickness of a twin structure period).

An interpretation of the supersonic (with respect to the longitudinal waves) growth rate of martensite crystals was obtained within the concept of a controlling wave process (CWP) [1, 2]. Given the supersonic growth rate of a martensite crystal, one needs a model of supersonic twinning. To this end, an important role was played by a reported [6] possibility of pumping up energy to the vibrating s-cells due to propagation of s-wave beams along mutually orthogonal directions $[110]_\gamma$ and $[1\bar{1}0]_\gamma$. This possibility, as for the first time shown in [7], allows one to understand (as early as when threshold strains are taken into consideration) how a regular transformation twin structure can be formed within a martensite crystal when only one initial active s-cell is available.

Moreover, in order to arrive at a complete description of a martensitic transformation, one needs a clear understanding of the transition from threshold strains, transferred by the controlling wave process and lying within the range of elastic deformations of the parent phase, to ultimate deformations exceeding the threshold ones by two – three orders. It is clear that only after the transition to ultimate strains one can unambiguously interpret the observed macroscopic morphological features as a macroshear and mutual orientation of the initial and final phase lattices. Note that the issue of transition from threshold to ultimate deformations was in principle solved in [8-12], using the bcc – hcp and fcc –bcc transformations as an example for the cases where the controlling wave process ensures the fastest rearrangement of the closest-packing of planes of the parent phase, disregarding crystal twinning. It is therefore necessary to consider the transition from threshold deformations to ultimate deformations for the case of martensite crystal twinning in the course of γ−α martensitic transformation as well, assuming that the controlling wave process stimulates Bain's deformation.

The answers to the questions posed above within the framework of the dynamic theory are reported in [13] published in Russian (with the contents translated into English given in Appendix. In this work, we present the most important results only.

## 2 Twin structure component ratio

The main component of a twin structure during initiation of Bain's deformation is associated with a dynamic cell shaped as a rectangular parallelepiped with its edges along three orthogonal axes $<001>_\gamma$. Two edges have equal dimensions

$$d_s < \lambda_s/2,$$

and the third dimension $d_{s\ell}$ satisfies the inequalities

$$d_s << d_{s\ell} < \lambda_\ell/2.$$

When a pair of s-wave beams propagating in the orthogonal directions $[100]_\gamma$ and $[0\bar{1}0]_\gamma$ overlaps, the growth speed of the main component of the twin structure in the $[1\bar{1}0]_\gamma$-direction by a factor of $\sqrt{2}$ exceeds the equal velocities $v_{1,2S}$ of the longitudinal s-waves in the $<001>_\gamma$ – directions

$$v_{tw} = v_{1,2S}\sqrt{2}.$$

These s-wave beams carry deformations that are equal in their value but different in their sign $\varepsilon_{1s} = -\varepsilon_{2s}$.

The relation for reproducibility of the best possible conditions for the short-wave cell to be activated in the region of the wave front of the controlling wave process is given by a simple formula

$$v_{1s} = v'_{2\ell}(\psi)\cos(\psi), \qquad (1)$$

where $v'_{2\ell}(\psi)$ is the projection of velocity $v_{2\ell}$ onto the plane $(001)_\gamma$, and $\psi$ is the angle between the direction of the projection of the wave normal $n_{2\ell}$ ($\ell$-wave carrying compressive deformation) and the $[100]_\gamma$ - axis in the $(001)_\gamma$ plane.

The values of $v_{2\ell}$ for the known elastic moduli of a material are found from the Kristoffel equations [14]. For the sake of illustration, Fig. 1 shows the distribution of the main (shaded areas) and additional (empty areas) components of the generating structure for $d_s = \lambda_s/4$ and $\psi = 26.6°$ for the following wave normal orientations:

$$\begin{aligned}n'_{1\ell} &= [0\ 0\ \bar{1}]_\gamma, \\ n'_{2\ell} &= [\cos\psi\ \sin\psi\ 0]_\gamma.\end{aligned} \qquad (2)$$

A characteristic feature of the structure is a strictly regular distribution of the components (the volume ratio in the case under study $\beta=2$).

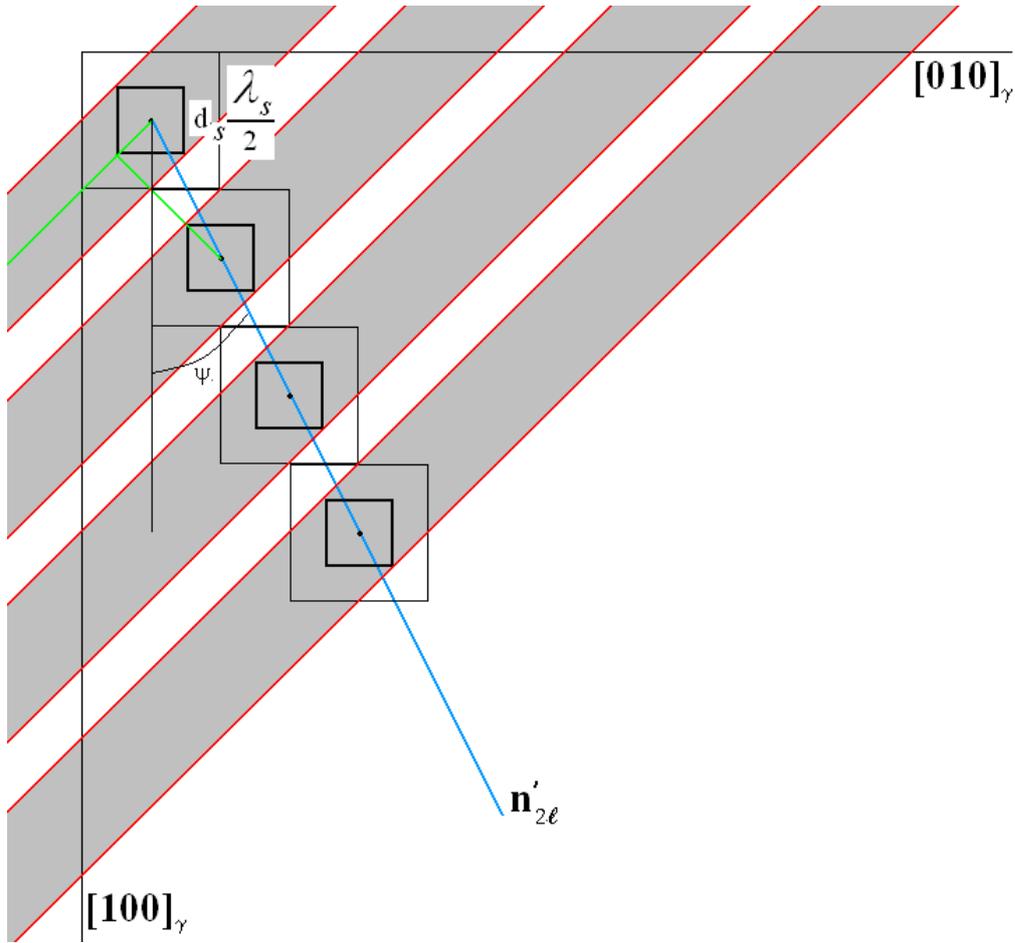

Fig. 1. Dynamic model of formation of a regular layered structure

The periodic layered structure being developed is characterized by the component volume ratio depending only on parameter $d_s/\lambda_s$ and direction $n_{2\ell}$ of propagation of the $\ell$-wave responsible for compressive deformation at the mesoscale level as part of the controlling wave process.

$$\beta = \frac{4\dfrac{d_s}{\lambda_s}}{1 + tg\psi - 4\dfrac{d_s}{\lambda_s}}. \qquad (3)$$

For the sake of illustration, Table 1 lists the reference values of $\beta$ and data sets of the parameters $tg\psi$ and $d_s/\lambda_s$, for which these $\beta$ values are achieved.

**Parameters** $\mathrm{tg}\psi$, $d_s/\lambda_s$ **and the corresponding values of** $\beta$

| tg$\psi$ | $\dfrac{d_s}{\lambda_s}$ | $\beta$ | tg$\psi$ | $\dfrac{d_s}{\lambda_s}$ | $\beta$ |
|---|---|---|---|---|---|
| 0 | $\dfrac{1}{4}$ | $\infty$ | $\dfrac{1}{3}$ | $\dfrac{2}{9}$ | 2 |
| 0 | $\dfrac{1}{6}$ | 2 | $\dfrac{1}{3}$ | $\dfrac{1}{5}$ | $\dfrac{3}{2}$ |
| 0 | $\dfrac{3}{20}$ | $\dfrac{3}{2}$ | $\dfrac{1}{3}$ | $\dfrac{1}{6}$ | 1 |
| 0 | $\dfrac{1}{8}$ | 1 | 1 | $\dfrac{1}{2}$ | $\infty$ |
| $\dfrac{1}{3}$ | $\dfrac{1}{3}$ | $\infty$ | 1 | $\dfrac{1}{3}$ | 2 |
| $\dfrac{1}{3}$ | $\dfrac{4}{15}$ | 4 | 1 | $\dfrac{3}{10}$ | $\dfrac{3}{2}$ |
| $\dfrac{1}{3}$ | $\dfrac{1}{4}$ | 3 | 1 | $\dfrac{1}{4}$ | 1 |

This result allows one to treat layered-structure formation (and twin martensite crystals in particular) as a purely dynamic process.

An optimal value of parameter $\beta = \beta_{tw}$ for a twin structure to be formed corresponds to the fastest lattice transformation in the course of martensitic transformation. Note that the tensile deformation $\varepsilon_{1B}$ in the main components of the regular layered structure bounding the additional component ensures the occurrence of Bain's compressive deformation $\varepsilon_{2B}$ in the additional component of the layered structure. The excessive compressive deformation developed in the orthogonal direction $\varepsilon_{1com} = |\varepsilon_{2B}| > \varepsilon_{1B}$ in the additional component of the regular layered structure corresponds in the first stage of deformation (within the time scale $T_s/2$, where $T_s$ is the period of s- oscillations) to the lacking compressive deformation in the main components. The deformation balance between the components of the twin structure is eventually established in the course of the second stage of deformation (within the time scale $T_\ell/2$, where $T_\ell$ is the period of the ℓ- oscillations) when compressive deformation in direction $\mathbf{n}_{2\ell}$ achieves its ultimate value $|\varepsilon_{2\ell}|_f$. Now we can show that

$$\beta = \beta_{tw} = \frac{|\varepsilon_{2B}|}{\varepsilon_{1B}}. \tag{4}$$

## 3  Summarized data on twinned $\{31015\}_\gamma$ – crystals

Table 2 presents summarized data on morphological features of twinned crystals. The results of calculations performed using the dynamic theory shown in the last column are given for the elastic modulus values (in TPa): $C_L = 0.2180$, $C' = 0.0270$, and $C^0{}_{44} = 0.1120$. These moduli, according to [15], correspond to the Fe-31.5%Ni alloy at the onset temperature of martensitic transformation $M_s$=239K. Since the experimental data on morphological properties refer to the Fe–22Ni–0.8C alloy, the moduli used provide a model set. Based on the calculation of the elastic field of a dislocational loop, the following wave normals were selected as directions:

$$\begin{aligned}\mathbf{n}_{1\ell} &= [0.147188 \quad -0.987456 \quad 0.057158]_\gamma, \\ \mathbf{n}_{2\ell} &= [0.948816 \quad 0.157282 \quad 0.273879]_\gamma.\end{aligned} \tag{5}$$

Using the Kristoffel's equation [14] for the normals (5) we obtain the speed ratio $æ_\ell = v_2/v_1 \approx 1.08861$, for which orientation of the normal with respect to an expected habitus

$$\mathbf{N}_{wl} \parallel [0.533462 \quad 0.14318 \quad 0.833617]_\gamma \qquad (6)$$

differs from $[10\ 3\ 15]_\gamma$ by 1.509°.

For the Fe–22Ni–0.8C alloy, during a γ–α martensitic transformation, according to [16], its relative volume variation and martensite tetragonality are δ=0.0384 and t=1.045, respectively. Hence, the main values of Bain's deformation would be

$$\varepsilon_{1,3B} \approx 0.12011, \quad \varepsilon_{2B} \approx -0.17232. \qquad (7)$$

According to Eqs. (4) and (7), the component ratio is as follows:

$$\beta = \beta_{tw} \approx 1.434685.$$

And the twinned component fraction is

$$\delta_{tw} = (\beta_{tw} + 1)^{-1} = \delta_{tw} \approx 0.41073. \qquad (8)$$

Since the normals (5) are deflected from the fourth-order axes of symmetry, the difference of quasi-longitudinal waves from purely longitudinal ones affects orientation of eigenvectors of the strain tensor, comparable to the ℓ-waves, and, therefore, also influences the results of calculation of orientation relationships and macroshear.

The asterisk (*) in Table 2 marks the data on macroshear from [17] (direction of macroshear is d=[–0.21017 –0.61850 0.75715]$_\gamma$, module is 0.19144) and from [18] (direction of macroshear is d=[–0.1761 –0.6886 0.7034]$_\gamma$, module is 0.1915), which we recalculated for the macroshear components along the habitus plane.

Table 2

**Summarized data on morphological features of twinned {3 10 15}$_\gamma$ crystals**

| Characteristic | Experiment | | Theoretical analysis | | |
|---|---|---|---|---|---|
| | G–T [16] | D–B [18] | V–L–R [19] | B–M [17] | This work |
| Habitus plane (HP) | 0.5472 0.1642 0.8208 | 0.5481 0.1748 0.8180 | 0.5691 0.1783 0.8027 | 0.55102 0.18726 0.81321 | 0.533462 0.14318 0.833617 |
| Angle between (HP)$_t$ and (HP)$_{G–T}$ | | | ≈1.728° | ≈1.284° | ≈1.509° |
| Orientation relationships | | | | | |
| $(111)_\gamma \wedge (101)_\alpha$ | ≈1° | ≈1° | ≈0.25° | ≈0.2° | ≈1.3108° |
| $[1\bar{1}0]_\gamma \wedge [11\bar{1}]_\alpha$ | ≈2.5° | ≈2.5° | ≈3° | ≈2.7° | ≈2.4319° |
| $[\bar{2}11]_\gamma \wedge [10\bar{1}]_\alpha$ | ≈2° | | ≈1.72° | ≈1.9° | ≈2.3089° |
| $[\bar{1}01]_\gamma \wedge [\bar{1}11]_\alpha$ | ≈6.5° | | ≈6.32° | ≈6.6° | ≈6.9810° |
| Angle of macroshear | 10.66° | * 10.69° | 10.33° 10.71° | * 10.5395° | 10.7839° |
| Direction of macroshear | −0.7315 −0.3828 0.5642 | * −0.7914 −0.2083 0.5747 | −0.7660 −0.2400 0.5964 | *−0.769969 −0.261645 0.581970 | −0.791386 −0.263371 0.551673 |
| $\Delta S \wedge \Delta S_{G-T}$ | | | ≈8.6° | *≈7.3702° | ≈7.7050° |
| $\Delta S \wedge \Delta S_{D-B}$ | | | *≈2.6081° | *≈4.2575° | ≈3.4243° |

Together with the results reported in [16] we included later experimental data [18] obtained for the same Fe–22Ni–0.8C alloy. Comparison of these results demonstrates that the direction of macroshear in [18] is much closer to the calculated data than in [16]. This refers both to the crystal geometry [17, 19] and to the present calculations.

## 4 Summary

A transition from threshold deformations, lying in the elastic region, to ultimate deformations and estimation of all morphological features observed (main-to-twinned component ratio in the twin structure, habitus plane orientation, orientation relationships and macroshear parameters), given their consistency with the experimental data, provide solid grounds to argue that the final stage of constructing the dynamic theory of martensitic transformations has been completed. This statement is valid as concerns, at least, the $\gamma-\alpha$ martensitic transformation in iron-based alloys with the grain size exceeding the critical one [20].

**Appendix**

**M.P. Kashchenko and V.G. Chashchina, Dynamic Model for Formation of Twinned Martensite Crystals upon γ–α Transformation in Iron-based Alloys** [in Russian], Ekaterinburg, The Ural State Forest Engineering University (2009), 98p.(http://www.nanometer.ru/2010/01/13/martensitnie_prevrashenia_162444.html)

CONTENTS

**Introduction**